\documentclass[runningheads.10pt]{llncs}
\usepackage{algorithmic}
\usepackage{graphicx}
\usepackage{epsfig}
\usepackage{color}
\usepackage{array}
\usepackage{amsmath}
\usepackage{fontenc}
\usepackage{textcomp}
\usepackage{rotating}
\usepackage{soul}
\usepackage{tabulary}
\usepackage{tabularx}
\usepackage{caption}
\usepackage{libertine}
\usepackage{url}
\usepackage{tikz}
\usepackage{balance}
\usepackage{amsfonts}
\usepackage{booktabs,pifont}
\usepackage{floatrow}
\usepackage{fancyvrb}
\usepackage{xcolor}
\usepackage{lipsum}%http://ctan.org/pkg/lipsum
\usepackage{array,multirow,graphicx}
\newtheorem{mydef}{Definition}
\newtheorem{mypro}{Property}

\usepackage[vlined,ruled,linesnumbered]{algorithm2e}
\usepackage{pdflscape,array,booktabs}
\makeatletter
\newcommand{\mathleft}{\@fleqntrue\@mathmargin0pt}
\newcommand{\mathcenter}{\@fleqnfalse}
\makeatother
\usepackage[export]{adjustbox}
\usepackage{caption, subcaption, floatrow}

\begin{document}
\title{FuncVul: An Effective Function Level Vulnerability Detection Model using LLM and Code Chunk}

% \title{Function Level Vulnerability Detection using Large Language Model}

\titlerunning{Function Level Vulnerability Detection using LLM}

\author{Sajal Halder \and
Muhammad Ejaz Ahmed  \and 
Seyit Camtepe %\inst{1}
}

\authorrunning{S. Halder et al.}
\institute{Data61, CSIRO, Australia \\  
\email{(sajal.halder,ejaz.ahmed,seyit.camtepe)@data61.csiro.au}}

\maketitle              % typeset the header of the contribution
\begin{abstract}
Software supply chain vulnerabilities arise when attackers exploit weaknesses by injecting vulnerable code into widely used packages or libraries within software repositories. While most existing approaches focus on identifying vulnerable packages or libraries, they often overlook the specific functions responsible for these vulnerabilities. Pinpointing vulnerable functions within packages or libraries is critical, as it can significantly reduce the risks associated with using open-source software. Identifying vulnerable patches is challenging because developers often submit code changes that are unrelated to vulnerability fixes. To address this issue, this paper introduces FuncVul, an innovative code chunk-based model for function-level vulnerability detection in C/C++ and Python, designed to identify multiple vulnerabilities within a function by focusing on smaller, critical code segments. To assess the model's effectiveness, we construct six code and generic code chunk based datasets using two approaches: (1) integrating patch information with large language models to label vulnerable samples and (2) leveraging large language models alone to detect vulnerabilities in function-level code. To design FuncVul vulnerability model, we utilise GraphCodeBERT fine tune model that captures both the syntactic and semantic aspects of code. Experimental results show that FuncVul outperforms existing state-of-the-art models, achieving an average accuracy of 87-92\% and an F1 score of 86-92\% across all datasets. Furthermore, we have demonstrated that our code-chunk-based FuncVul model improves 53.9\% accuracy and 42.0\% F1-score than the full function-based vulnerability prediction. The FuncVul code and datasets are publicly available on GitHub \footnote{\url{https://github.com/sajalhalder/FuncVul}. \newline In The 30th European Symposium on Research in Computer Security (ESORICS), 22 Sep - 26 Sep, 2025, Toulouse, France. }

\end{abstract}

\section{Introduction} 

With the rapid expansion of technology, cybersecurity has become a growing priority. By October 2024, the National Vulnerability Database (NVD) recorded over 240,000 reported Common Vulnerabilities and Exposures (CVEs) \cite{CVE2024,NVD}. This number has steadily risen, with an average growth rate of 15-20\% per year. Detecting vulnerabilities in C/C++ and Python code is a challenging process that demands thorough analysis of the codebase's structure, syntax, and semantics to reveal potential weaknesses exploitable by attackers. Identifying vulnerable functions within the package is crucial because it allows developers to focus their efforts on resolving the specific issue rather than discarding the entire package. It enables organizations to prioritize fixes based on the severity and significance of the affected functions. Moreover, identifying vulnerable source enables the developer to resolve them quickly, minimizing the impact on customers and ensuring service continuity. %Besides that, after identifying the vulnerable issues, the developer can resolve them quickly, which minimizing the impact on customers and maintaining service continuity. 
Identifying vulnerable functions within the vast number of packages released daily is both time-consuming and requires specialized expertise in security. Thus, an automated model capable of effectively identifying vulnerable functions is essential. %that can identify vulnerable functions effectively is very essential.   

Existing research has explored code similarity techniques to detect vulnerable code patterns using machine learning \cite{sonnekalb2019machine}, deep learning \cite{chakraborty2021deep,wartschinski2022vudenc,li2021vuldeelocator}, and graph-based models \cite{wang2020combining}. Yuan et al. \cite{yuan2024software} combined serialized features from Gated Recurrent Units (GRUs) and structural features from Abstract Syntax Trees (ASTs) via Gated Graph Recurrent Networks (GGRNs), addressing data scarcity and imbalance with a Random Forest model, achieving superior performance. Vo et al. \cite{vo2023can} found that pre-trained deep models for vulnerability type identification (VTI) offered limited improvement over classical TF-IDF baselines and enhanced them by identifying key code tokens. Wang et al. \cite{wang2024reposvul} developed ReposVul, a repository-level dataset created using an automated framework with modules for untangling vulnerabilities, dependency extraction, and filtering outdated patches. Other works explored context-aware embeddings \cite{wei2021context}, RoBERTa models pre-trained on open-source C/C++ code \cite{hanif2022vulberta}, and Word2Vec-LSTM pipelines for Python code \cite{wartschinski2022vudenc}.  However, these approaches face two key limitations. First, while they can identify whether a function is vulnerable, they cannot determine the exact number of vulnerabilities within the function. Second, pinpointing the precise lines of code that contain vulnerabilities remains a challenge. As a result, security analysts are compelled to manually review functions, significantly increasing the time and effort required for vulnerability analysis. Addressing these limitations is vital for streamlining the vulnerability identification process and improving efficiency.

To address these limitations, we propose a code chunk-based function vulnerability detection model capable of identifying multiple vulnerabilities within a function. Moreover, it highlights smaller code chunks containing vulnerabilities, significantly reducing the time required for experts to address the issues. To sum up, in this paper we aim to answer the following research questions.

\begin{itemize}
\item [RQ1:] What modeling strategies can be employed to accurately detect function level vulnerabilities? 

\item [RQ2:] Does leveraging code chunks enhance model performance compared to analyzing full-function code?

\item [RQ3:] Does the FuncVul model leverage generalized code properties for vulnerability detection? 

\item [RQ4:] How effective is our approach at detecting vulnerabilities in unseen projects? 

\item [RQ5:] How does the performance of our approach vary with different numbers of source lines in a code chunk? 

\item [RQ6.] Is our proposed model capable of detecting multiple vulnerabilities within a single function's code?
\end{itemize}

To evaluate the performance of our proposed FuncVul model, we compared it against several state-of-the-art models: CodeBERT \cite{feng2020codebert}, CustomVulBERTa \cite{hanif2022vulberta}, BERT \cite{kenton2019bert} and VUDENC \cite{wartschinski2022vudenc} across six datasets. The main contribution of this research work are as follows.

 \begin{itemize}
 \item We propose a novel code chunk-based \textbf{\underline{Func}}tion \underline{\textbf{Vul}}nerability (\textbf{FuncVul}) detection model capable of identifying multiple vulnerabilities within a function and, importantly, to identify the specific, smaller code segments responsible for those vulnerabilities.
\item We collected and curated four datasets from diverse data sources, such as project source codes from GitHub and vulnerability advisory databases, i.e., OSV. Additionally, we developed novel methods to analyze, process, and curate source code using large language models (LLMs). 

\item We employ a fine-tuned GraphCodeBERT model for function-level vulnerability prediction, as it effectively captures both syntactic and semantic similarities within the code.
\item Our experimental results demonstrate that the proposed FuncVul model outperforms state-of-the-art baselines, achieving an average accuracy of 89.39\% and an F1 score of 88.94\% across the six datasets.
\item Additionally, we show that the FuncVul model is highly generic, capable of handling diverse code chunks and identifying new vulnerable patterns effectively.

 \end{itemize}

The remaining part of the paper is organized as follows. We briefly describe the relevant existing works in Section~\ref{existingwork}. Then, we discuss the problem statement in Section~\ref{problem}. We introduce our proposed model in Section~\ref{model}. After that, we present our experiments in Section~\ref{experiments}. Finally, we conclude the paper with potential future research directions in Section~\ref{conclusion}.

 \section{Existing Works}
\label{existingwork}
Software vulnerability detection is a significant challenge for security researchers in both academia and industry. Researcher classified vulnerability research in two directions: vulnerability dataset creation and vulnerability sections. This section reviews related work on function label vulnerability detection.

Hanif et al. \cite{hanif2022vulberta} presented a deep learning framework named VulBERTa to identify security vulnerabilities in source code. It leverages a RoBERTa model pre-trained with a specialized tokenization pipeline on real-world open-source C/C++ code considering code syntax and semantics. Warschinski et al. \cite{wartschinski2022vudenc} proposed VUDENC, a deep learning model for detecting vulnerabilities in Python code, combining Word2Vec embeddings with an LSTM network to classify vulnerable code sequences. Yuan et al.\cite{yuan2024software} proposed a hybrid approach combining GRU-based serialized features and Gated Graph Recurrent Network (GGRN0 based structural features, using a Random Forest model to improve vulnerability identification under data scarcity and imbalance. Vo et al. \cite{vo2023can} showed that deep pre-trained models offer limited gains over a TF-IDF baseline for vulnerability type identification and proposed a lightweight enhancement to identify key tokens for each vulnerability type.

Wang et al. \cite{wang2024reposvul} developed ReposVul, a repository-level vulnerability dataset built using an automated framework with modules for untangling fixes, extracting multi-level dependencies, and filtering outdated patches. Wei et al. \cite{wei2021context} proposed a supervised framework that utilised pre-trained context-aware embeddings (ELMo) and a Bi-LSTM layer to capture deep contextual representations and learn long-range code dependencies in source code to detect function vulnerability. Li et al. proposed $VulPecker$ \cite{li2016vulpecker}, an automated tool to identify known vulnerabilities within software source code. It utilises features derived from patches and employs a variety of code-similarity algorithms. Fu et al. \cite{fu2022linevul} designed a Transformer-based line-level vulnerability prediction method named $LineVul$ to detect vulnerabilities in C/C++ codes.  Tran et al. \cite{tran2025detectvul} implement DetectVul which is a statement-level vulnerability detection approach for Python that uses self-attention to learn patterns directly from raw code, avoiding graph extraction. Li et al.\cite{li2018vuldeepecker} proposed \emph{VulDeePecker} a code gadget-based model that represents programs as vectors by grouping semantically related, though not necessarily consecutive, lines of code.

GNN-based methods have demonstrated state-of-the-art performance in vulnerability detection by leveraging graph representations of source code. Hin et al. \cite{hin2022linevd} introduced a deep learning framework called $LineVD$ for statement-level vulnerability detection in C/C++ codes that combines GNNs and transformers. Notable approaches include Devign \cite{zhou2019devign}, which uses graph-level classification with semantic code representations, and GGNN-based methods \cite{wang2020dynamic} that capture data, control, and call dependencies with majority voting from traditional classifiers. Li et al. \cite{li2021vulnerability} introduced a feature-attentive GCN on program dependency graphs, while VulCNN \cite{wu2022vulcnn} transformed source code into semantic-preserving images. ReGVD \cite{nguyen2022regvd} utilised token embeddings from pre-trained models, residual connections, and pooling techniques to enhance graph representations. AMPLE \cite{wen2023vulnerability} improved performance by refining graph structures and capturing distant node relations. Islam et al. \cite{islam2023unbiased} introduced Poacher Flow edges to bridge static and dynamic analyses and manage long-range dependencies for richer vulnerability detection.

Large pre-trained language models like BERT and GPT have become a dominant learning paradigm, achieving notable success in computer vision and NLP by leveraging semantic knowledge from large-scale corpora. This pre-trained and fine-tuned approach has also been extended to code-related tasks with models like RoBERTa \cite{liu2019roberta}, CodeBERT \cite{feng2020codebert}, and GraphCodeBERT \cite{guo2020graphcodebert}, significantly improving applications such as automated program repair \cite{xia2023automated} and code vulnerability detection. All these works primarily focus on either software package vulnerabilities or function-level vulnerabilities. Although the LineVul \cite{fu2022linevul} model predicts vulnerabilities at the line level, it may still miss certain vulnerabilities due to overlooked semantic patterns.  They fail to identify the number of vulnerable lines within the function code or pinpoint the exact location of the vulnerabilities. Therefore, research that can effectively address function-level vulnerabilities is critically needed. 

Now a days, LLM are used in different domains, including vulnerability detection. Lu et al. \cite{lu2024grace} proposed GRACE, a vulnerability detection framework that enhances LLM-based analysis by incorporating code's graph structural information and in-context learning. Akuthota et al. \cite{akuthota2023vulnerability} utilised LLM for the purpose of identifying and monitoring vulnerabilities. Meanwhile, Guo et al. \cite{guo2024outside} explored the ability of LLMs to detect vulnerabilities in source code by evaluating models outside their typical uses to assess their potential in cybersecurity tasks. However, our proposed model is different than the existing models, where we use LLM to generate datasets and use code based fine tune model to detect vulnerability.

\subsection{Differences with Previous Works} 

Our proposed function-level vulnerability detection model introduces several key advancements over state-of-the-art techniques. First, unlike existing approaches that analyse entire functions or line-based vulnerability detection, our model focuses on code chunk-based vulnerability detection in C/C++ and Python. This approach significantly reduces the time required by experts or developers to address vulnerabilities. Second, our code chunk-based method enables the detection of multiple vulnerabilities within a function, whereas existing models typically provide only a binary assessment of whether a function is vulnerable. Third, our model leverages a large language model that is capable of supporting code chunks from different programming languages, eliminating the need for language-specific preprocessing required by existing methods. Finally, we utilise the pre-trained GraphCodeBERT model to build a function-level vulnerability detection framework that effectively captures both syntactic and semantic features, surpassing traditional approaches that rely solely on code similarity.

\section{Preliminaries \& Problem Statement}
\label{problem}

In this section, we first present the key preliminary definitions and then describe the problem statement.

\begin{mydef}
[Function Code Chunk]: 
A Function Code Chunk (FC) refers to a contiguous segment of lines extracted from a function’s source code, typically centered around a code change or edit. It includes a few lines before and after the change to preserve local context for vulnerability analysis.  
\end{mydef}

\begin{mydef} [Generic Code Chunk]: 
Generic Code Chunk represents the segments of code where variable names, function names, and other identifiers have been replaced with generic placeholders (e.g., $F_1$, $F_2$,...,$F_n$ for functions and $V_1$, $V_2$, ..., $V_n$ for variables). 
\end{mydef}
This generic code chunk transformation standardizes the code, removing specific naming conventions or contextual biases, and ensures a consistent format that focuses on structural and syntactic patterns.

\begin{mydef} [3-Line Extended-Based Code Chunk]:
3-Line Extended-Based Code Chunk refers to a segment of code centered on an edited line (or lines), augmented with three preceding and three succeeding lines from the edited lines in the function. 
\end{mydef}

This design captures the semantic context of code for vulnerability detection. Generally, the code edited lines is fewer than 10 lines. If the edited length is more than 10 lines, we consider edited lines only to make the code chunk. Thus, we can define the code chunk as follows. 
\begin{equation}
\small
    \text{Code Chunk} = \begin{cases} 
\{ L_i \mid i \in [\min(E)-3, \max(E)+3] \} & \text{if } |E| \leq 10 \\
E & \text{if } |E| > 10
\end{cases}
\end{equation}
where $L_i$ represents the $i^{th}$ line of the function code, $min(E)$ and $max(E)$ refer starting and ending edited lines, respectively. This code chunk approach provides a contextualized view of the code, enabling better understanding and analysis of the detected lines within their surrounding context.

\textbf{Problem Definition:} Given a C/C++ or Python based software patch information based modified function code chunk $(fc_i)$. The main goal of this research work is to develop a vulnerable code detector $\mathcal{V}$ which can identify patch-modified codes as vulnerable or non-vulnerable. It can be defined as follows.

\begin{equation}
\small
    \mathcal{V}(fc_i)=
\begin{cases}
1,\quad \text{if $fc_i$ is vulnerable,}\\
0, \quad \text{non-vulnerable}
\end{cases}
\end{equation}

To solve the problem, we propose 3-line extended based code chunk to detect function label vulnerability using code-based fine-tune models in C/C++ or Python code.

\section{Proposed Model}
\label{model}

In this paper, we propose an effective            
                    function-level vulnerability detection framework that leverages large language models (LLMs) alongside specialized code vulnerability detection techniques. To generate ground truth data, we utilise two distinct types of LLM prompts and employ an additional prompt to transform code chunks into generic code chunks. Subsequently, we fine-tune the prediction models using advanced code vulnerability identification techniques. The next two subsections provide a detailed explanation of the data generation process and the proposed \emph{FuncVul} models.

\subsection{Data Generation}
\label{data_generation}
\begin{figure*}[htp!]
    \centering
     
    \includegraphics[width=1.0\linewidth]{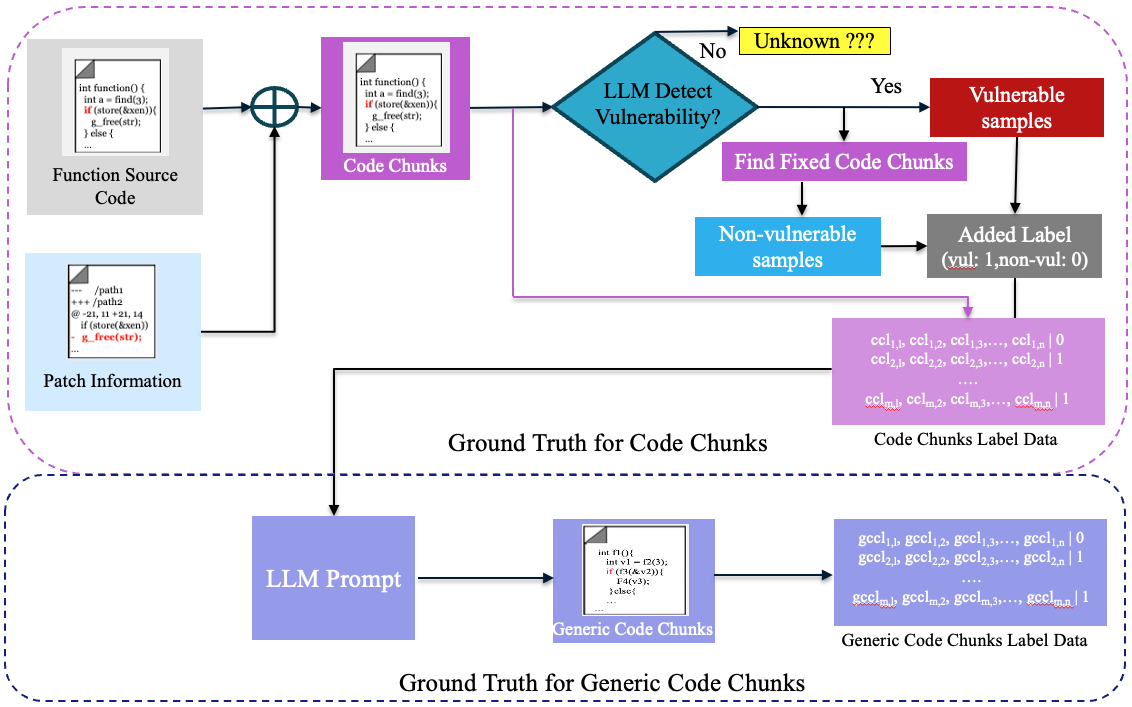}
    \caption{Code Checks and Generic Code Chunks Label Data Generation.}
    \label{fig:label_data}
\end{figure*}

Labeling data is critical for training any prediction model, yet identifying vulnerable data often poses significant challenges. In this study, we construct two types of ground truth datasets: code chunks and generic code chunks, derived from function source code and corresponding patch information. Figure \ref{fig:label_data} illustrates the processes involved in generating these datasets. Detailed descriptions of the ground truth generation for both code chunks and generic code chunks are provided in the following subsections.

In this study, we focus on function code chunks rather than full function code for two key reasons. First, vulnerabilities often exist within just one or two lines of code inside a function, and models trained on entire functions may struggle to pinpoint these specific vulnerable lines that potentially leading to inaccurate predictions. Second, using code chunks reduces the search space and minimizes the number of tokens processed by the tokenizer, enabling the fine-tuned model to more effectively distinguish between vulnerable and non-vulnerable patterns.

In this work, we generate code chunks by leveraging function source code and patch information. Patch information highlights the modifications made to the code, marking added lines with a plus sign (+) and removed lines with a minus sign (-) at the beginning of each line. Additionally, it includes a chunk header that specifies the location and range of the changes, indicating where the modifications begin and the consequences of changes using line numbers.

The algorithm~\ref{algo:code_chunk} extracts relevant code chunk segments from a function based on patch information. It first parses the patch details to retrieve the chunk header, removed lines, and added lines in line 1. Next, it extracts the starting lines and corresponding modification ranges for both the removed and added lines in line 2. Context parameters are initialized to include three lines before and after the modified region (for three line strategy) in line 3. The algorithm then initialises indices of the removed lines within the function, recording them in a list called modified\_index in line 4. If no matches are found, it returns an empty result in line 9. For matched lines, the algorithm determines the bounds of the code chunk using a heuristic for small regions ($\leq$10 lines) to include additional context, or directly uses the minimum and maximum indices for larger regions in lines 11-16. Finally, it extracts and returns the code chunk based on the calculated bounds in line 16 and line 17, respectively.
\begin{algorithm}[htp!]
\small 
\SetKwFunction{This}{this}
\KwData{ \textbf{F}: Function source code; \textbf{P}: Patch information.}

\KwResult{\textbf{FC}: Extracted function code chunk.}

chunk\_header, removed\_lines, added\_lines $\leftarrow$ \textbf{P}

removed\_start\_line, removed\_line\_range, added\_start\_line, added\_line\_range $\leftarrow$ chunk\_header 

before\_lines,  after\_lines = 3, 3

modified\_index $\leftarrow$ $\{\}$

\For { index, line $\in$ enumerate(\textbf{F}[removed\_start\_line : removed\_start\_line + removed\_line\_range])} 
{

    \If{ line $\in$ removed\_lines} 
    {
    
    modified\_index.append(removed\_start\_line + index)
    
    }

}

\If{modified\_index == $\{\}$ }{
    \Return $\{\}$
}

\eIf{max(modified\_index) - min(modified\_index) $\leq$ 10}{

    start\_index $\leftarrow$ max(modified\_index[0]) - before\_lines, 0)
    
    end\_index $\leftarrow$min(max(modified\_index[-1]) + after\_lines + 1, len(F))
    
}
{
    start\_index $\leftarrow$ min(modified\_index)
    
    end\_index $\leftarrow$ max(modified\_index)
}

Extract function code chunk FC = F[start\_index:end\_index]  

\Return \texttt{FC}\;
\caption{Find Function Code Chunk (F, P) }
\label{algo:code_chunk}
\end{algorithm}

\subsubsection{Generic Code Chunks: }

Generic code chunk converts code chunks to a generic format.  In this work, we have transformed function code into a generic format by renaming functions as $F_1$, $F_2$, ..., $F_m$ and variables as $v_1$, $v_2$, ..., $v_n$. The key advantage lies in mitigating the variations introduced by different developers who often use diverse functions and variable names to achieve the same functionality.  The LLM prompt designed to standardize code chunks by converting them into their generic format is presented in Appendix \ref{generic_prompts}.

In this work, we leverage the Gemini 1.5 Pro \cite{team2024gemini}  LLM model to efficiently transform generic code chunks. A key advantage of utilizing this LLM model is its ability to seamlessly convert code across various programming languages, including C/C++, Java, and Python.

\subsubsection{Vulnerable and Non-Vulnerable Samples:}
Our primary objective is to detect vulnerabilities using function-level code, whether in its original form as code chunks or transformed into generic code chunks. To develop a robust vulnerability detection model, we require a ground truth dataset comprising both vulnerable and non-vulnerable samples. To construct this dataset, we adopt a dual-strategy approach that combines code-based heuristics with predictions from a LLM. This methodology enhances the ability to identify functions with a higher likelihood of containing vulnerabilities, ensuring greater confidence in the dataset's accuracy.

\begin{mypro} 
\textbf{Code-Based Heuristic (Patch Modification Hypothesis):} We hypothesize that functions containing only a single modification within a CVE patch are more likely to contain the vulnerability. This hypothesis stems from the assumption that smaller, localized patches often address specific vulnerabilities directly. This property does not guarantee that the code chunks will always be vulnerable, as developers may modify patches to enhance code quality.
\end{mypro}

Our study consists of clean and localized vulnerability cases from OSV.dev, where our empirical study shows 80.04\% (6515 out of 8139) of CVEs have a single Git commit patch. Therefore, we restricted our study to single-patch CVEs, aligning with VFCFinder \cite{dunlap2024vfcfinder}. Multiple modifications make it unclear which change corresponds to the vulnerability. For dataset reliability, we excluded multifile patch information. Each modified patch contains chunk headers with deleted and added lines between code versions. The before version shows the vulnerable state, while the after version shows the fixed code.

\begin{mypro} 
\textbf{LLM-Based Heuristic (Vulnerable Line Detection):} We utilise a LLM Gemini-1.5 Pro \cite{team2024gemini} to predict vulnerable lines within code chunks. This model is presented with either (i) the code chunk alone or (ii) the code chunk alongside its corresponding CVE description. This dual input strategy aims to leverage both code structure and vulnerability context for improved ground vulnerability predictions. Appendix \ref{vulnerable_samples_prompts} shows the two different prompts that we use in this work to identify vulnerable samples. 

\end{mypro}
\emph{\textbf{Vulnerable Ground Truth:}} A code chunk is classified as vulnerable (class label: 1) and included in the ground truth dataset if it satisfies the following criteria:

\begin{itemize} 
\item \textbf{Property 1} must be fulfilled.
\item According to \textbf{Property 2}, the LLM response for \textit{vul\_lines} is not \textit{None}. 
\item There is at least one overlapping line between the \textit{vul\_lines} identified by the LLM and the deleted lines in the patch modification.
\end{itemize}

If any of the above criteria are not met, the code chunk is labeled as Unknown (see Figure \ref{fig:label_data}).

\emph{\textbf{Non-Vulnerable Ground Truth:}} 
After the labeling of vulnerable code chunks, we extract fixed code from the after version using patch modification details and construct non-vulnerable code chunk samples. Additionally, we include random 5 to 10 lines of code from fixed functions in the after version. These samples are classified as non-vulnerable (class label: 0).

\subsubsection{Code Chunks and Generic Code Chunks Label Data:}

Figure \ref{fig:label_data} illustrates the process of generating labeled data for code chunks and generic code chunks. These chunks are constructed using two types of LLM prompts (detailed in Table \ref{prompt_1_2}). Therefore, based on two LLM prompts and code chunks and generic code chunks, we generate four label datasets (Dataset 1, Dataset 2, Dataset 3 and Dataset 4) that shown in dataset section (\emph{c.f.} Section \ref{dataset_sec}) in Table \ref{dataset}. 

We further created two additional datasets, Dataset 5 and Dataset 6 (\emph{c.f.} Section \ref{dataset_sec}) , by providing the full function code to a large language model (LLM) to identify vulnerable lines. If the LLM successfully detects at least one vulnerable line, we apply the $N$-line code chunking approach to generate positive samples. The same strategy used for generating negative samples in Datasets 1–4 is applied here for consistency.

\subsection{Proposed FuncVul Model}

\begin{figure}
    \centering
    \includegraphics[width=1.0\linewidth]{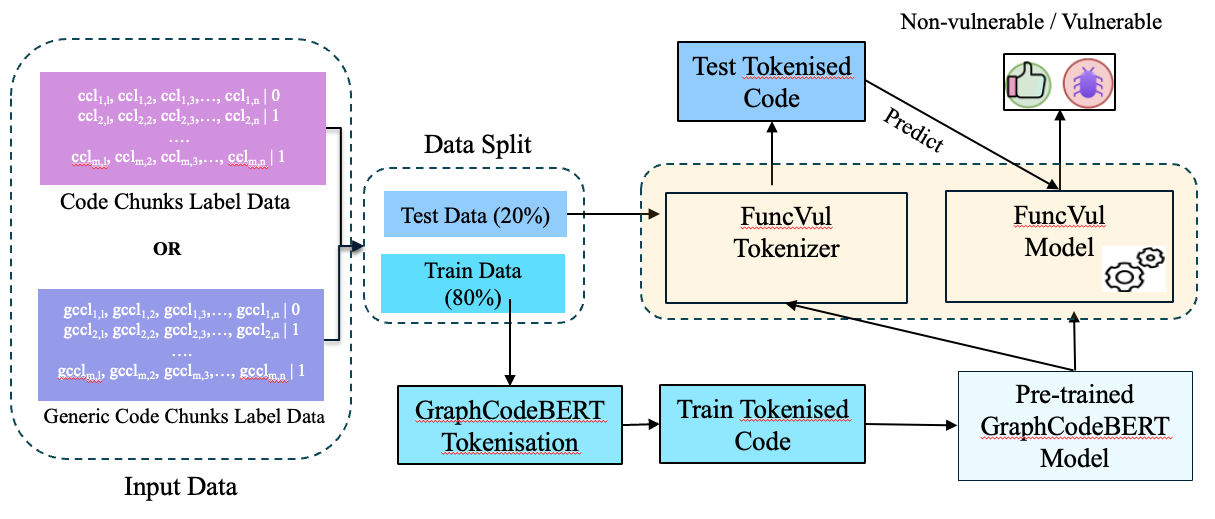}
    \caption{Proposed code chunk based function vulnerability detection model (FuncVul) architecture.}
    \label{architecture}
\end{figure}

Figure \ref{architecture} provides an overview of the architecture for the proposed FuncVul model, designed to detect function-level vulnerabilities effectively. The process begins with the input data, which consists of either code chunks or generic code chunks. These inputs are preprocessed and split into two subsets: 80\% for training and 20\% for testing. The training data is then tokenized using the tokenizer from the pre-trained GraphCodeBERT model. This tokenization step transforms the raw code chunks into numerical representations that encode the syntactic and semantic features of the code. Subsequently, the tokenized training data is passed through the pre-trained GraphCodeBERT model, which has been fine-tuned to capture rich features specific to programming languages. 

GraphCodeBERT \cite{guo2020graphcodebert}  is a pre-trained model for programming languages that incorporates the semantic structure of code, focusing on data flow rather than abstract syntax trees (AST). Data flow represents variable relationships through a graph, simplifying complexity and enhancing efficiency. The model introduces two structure-aware pre-training tasks: data flow edge prediction to learn code structure representation and variable alignment to bridge source code and data flow representations. Built on a Transformer \cite{vaswani2017attention} architecture, GraphCodeBERT extends it with a graph-guided masked attention mechanism, enabling it to effectively integrate code structure and improves code representation learning.

In Figure 2, we illustrate the fine-tuning process of the GraphCodeBERT model using our generated code chunk or generic code chunk data. After fine-tuning, the GraphCodeBERT model builds a new model and tokeniser, which we refer to as the FuncVul model and FuncVul tokeniser, respectively. To evaluate the model’s performance on the test data, the test code is first tokenised using the FuncVul tokeniser. The tokenised code is then fed into the FuncVul model, which predicts whether the code chunk is vulnerable or non-vulnerable.

% \subsubsection{Gemini Fine Tune Model}
%  Gemini 1.5 Flash \cite{team2024gemini} are advanced multimodal models designed for exceptional efficiency, reasoning, planning, multilingual capabilities, and long-context performance. Leveraging innovations in sparse and dense scaling, training, distillation, and infrastructure, these models excel at processing extensive contexts of up to 10M tokens.
% % \subsubsection{OpenAI Fine Tune Model}

\subsection{FuncVul algorithm}

\begin{algorithm}[htp!]

\SetKwFunction{This}{this}
\KwData{ Data: Code chunks or Generic code chunks data}

\KwResult{FuncVul model $\mathcal{M}$ and FuncVul tokeniser $\mathcal{T}$}

\textbf{Training Phase:}

Split Data into train\_data (80\%) and test\_data (20\%)

$\mathcal{M_G}$, $\mathcal{B_G}$ $\leftarrow$ Load GraphCodeBERT tokeniser and model 

train\_tokenised\_code $\leftarrow$ Tokenize train\_data using GraphCodeBERT tokenizer $\mathcal{T_G}$(test\_data).

Set training parameters (e.g., epochs, batch size, logging steps, learning rate).

$\mathcal{M}$, $\mathcal{B}$ $\leftarrow$ Train GraphCodeBERT using train\_data $\mathcal{M_G}$(train\_tokenised\_code) and parameters and generate new model and tokeniser 

Save the FuncVul model $\mathcal{M}$ and FuncVul tokenizer. 

\textbf{Testing Phase:}

test\_tokenised\_code $\leftarrow$ Tokenize test\_data using FuncVul tokenizer $\mathcal{T}$(test\_data).

predict\_label  $\leftarrow$ Predict vulnerabilities (vulnerable or non-vulnerable) using FuncVul model $\mathcal{M}$ (test\_tokenised\_code).

Evaluate the model's performance on the test\_data label and predict\_label.

\Return \texttt{$\mathcal{M}$, $\mathcal{T}$}
\caption{FuncVul Model (Data)}
\label{algo:fine_tune_model}

\end{algorithm}

The algorithm \ref{algo:fine_tune_model} presents the training and testing process for the function-level vulnerability detection model, FuncVul. During the training phase, the data is split into training (80\%) and testing (20\%) sets (line 2). Next, the GraphCodeBERT tokenizer and model are loaded (line 3), and the training data is tokenized using the GraphCodeBERT tokenizer to create train\_tokenised\_code (line 4). Training parameters, such as epochs, batch size, and learning rate, are configured (line 5). Using the tokenized training data, the GraphCodeBERT model is fine-tuned to produce the FuncVul model $\mathcal{M}$ and tokenizer $\mathcal{T}$ (line 6). These are saved for future use (line 7).

In the testing phase, the test data is tokenized using the FuncVul tokenizer $\mathcal{T}$ (line 9). The FuncVul model $\mathcal{M}$ then predicts vulnerability labels for the tokenized test data (line 10). Model performance is evaluated by comparing the predicted labels with the ground truth labels from the test data (line 11). Finally, the algorithm returns the FuncVul model and tokenizer (line 12).

\section{Experiments}
\label{experiments}

\subsection{Experimental Setup} 
All experiments in this paper were conducted using Python on a MacBook Pro with an Apple M3 processor and 24GB of RAM. For the FuncVul implementation, a batch size of 8 was used with a chunk code embedding vector length of 512. The model was trained for 3 epochs with 50 warmup steps, a weight decay of 0.05, and automatic reloading of the best model at the end.

\subsubsection{Datasets:}
\label{dataset_sec}

In this research work, we generate six datasets. The dataset generation process is discussed in detail in the corresponding section (\emph{c.f.} Section \ref{data_generation}). The first four datasets were created by combining patch information (removed lines) with lines detected by the LLM, ensuring that at least one common line is present between the two. In contrast, datasets 5 and 6 were generated solely using code chunks identified by the LLM, without incorporating any removed line information. For identifying vulnerable lines using LLMs, we employ two prompts: one that utilises only the code information and another that incorporates both the code and its description.  Table \ref{dataset} provides details of six datasets, including the code type, and the number of vulnerable and non-vulnerable samples. 
\begin{table}[htp!]
\centering
\caption{Details of various datasets.   }
\def\arraystretch{1.0}\tabcolsep=3pt
\resizebox{0.95\textwidth}{!}{
\begin{tabular}{c l   l  l c  c   }
\hline

 Dataset & Prompt & Code Type & Vulnerable Defined By & Vulnerable &  Non-vulnerable \\ 
% &&  &  & &  \\
\midrule 

1 &  Code + Description & Code Chunk & LLM + Patch Information & 1810 (43.4\%) & 2357 (56.6\%)
\\

\midrule 
2 & Code & Code Chunk & LLM + Patch Information &  2120 (42.6\%) & 2851 (57.4\%) \\ \midrule

3 &  Code + Description & Generic Code Chunk &  LLM + Patch Information & 1810 (43.4\%) & 2357 (56.6\%)
\\ \midrule

4 & Code & Generic Code Chunk & LLM + Patch Information & 2120 (42.6\%) & 2851 (57.4\%) \\ \midrule

5 & Code + Description & Code Chunk & LLM  &  3169 (50\%) &  3169 (50\%) \\ \midrule

6 & Code & Code Chunk & LLM  & 6041 (50\%) & 6041 (50\%) \\ \bottomrule

\end{tabular}
 }
\label{dataset}
\end{table}

\subsubsection{Baselines:}

We compare our proposed FuncVul model with five baselines: \textbf{CodeBERT} \cite{feng2020codebert}, \textbf{CustomVulBERTa} \cite{hanif2022vulberta}, \textbf{BERT} \cite{kenton2019bert}, \textbf{VUDENC} \cite{wartschinski2022vudenc} and \textbf{LineVul} \cite{fu2022linevul}. These baselines' detailed descriptions are given in Appendix \ref{baselines}.

\subsubsection{Evaluation Metrics:}

In the prediction models analyses, we applied various evaluation metrics indicating the model performances. Our main goal is to predict code vulnerability. Thus, we evaluate our results using Accuracy, Precision, Recall, F1-score and Matthews Correlation Coefficient (MCC). The details of these evaluation metrics are defined in Appendix \ref{evaluatoin_metric_details}.

\subsection{Results Analysis}
To evaluate the proposed model FuncVul performance, we run a set of experiments to answer our six research questions. 
\begin{table}[h!]
\small
    \centering
    \caption{Comparison of FuncVul and baselines across six datasets, with bold for best scores, underline for second-best, and bracketed numbers indicating F1-score rankings (1 = best, 6 = worst).}

    \resizebox{0.98\textwidth}{!}{
   \begin{tabular}{c|lcccc c }
\toprule 
 Dataset  & Model & Accuracy & Precision & Recall & F1-Score & MCC \\
\midrule
  &   CodeBERT  & 0.8707±0.00020 & 0.7928±0.0216 & 0.9505±0.0269 & 0.8641±0.0095 (5) & 0.7683±0.0283\\

 % & Gemini& 0.6310±0.0902 &      0.6422±0.1286 &  0.3137±0.2162 &  0.4018±0.2223 (6)\\ 
1 & CustomVulBERTa & 0.8648±0.0060 & 0.7974±0.0224 & \underline{0.9816±0.204} & 0.8816±0.0091 (3) & 0.7898±0.0102\\ 
 & BERT & 0.8812±0.0103 & \underline{0.8067±0.0168} & 0.9548±0.0443 & 0.8739±0.0167 (4) & 0.7742±0.0282\\
 & VUDENC & 0.8598±0.0111 & 0.8058±0.0090 & 0.8560±0.0380 & 0.8404±0.0225 (6) & 0.7166±0.0241 \\
 & LineVul & \underline{0.8874±0.0074} & 0.7937±0.0174 & 0.9802±0.0.0 & \underline{0.8849±0.0708} (2) & \underline{0.7976±0.0118} \\
 &  \textbf{FuncVul}   & \textbf{0.8906±0.0042} & \textbf{0.8108±0.0136} & \textbf{0.9840±0.0206} & \textbf{0.8888±0.0055} (1) & \textbf{0.9477±0.0025}\\
 \midrule

 & CodeBERT  & \underline{0.8950±0.0117} & \underline{0.8151±0.0315} & 0.9777±0.0322 & \underline{0.8882±0.0111}  (2) & \underline{0.8039±0.0183} \\

  % & Gemini & 0.6253±0.1002 &     0.6431±0.1266 &  0.3636±0.2209 &   0.4314±0.2441 (6) \\ 
 2 & CustomVulBERTa & 0.8908±0.0130 & 0.7975±0.0232 & \underline{0.9976±0.0053} & 0.8863±0.0132 (3) & 0.8022±0.0200\\
  & BERT & 0.8902±0.0119 & 0.8136±0.0241 & 0.9650±0.0361 & 0.8822±0.0126 (5) & 0.7919±0.0244 \\ 
  & VUDENC & 0.8680±0.0140 & 0.8463±0.0183 & 0.8440±0.0301 & 0.8449±0.0183 (6) & 0.7304±0.0295\\
  & LineVul & 0.8900±0.0120 & 0.7951±0.0202 & \textbf{1.0±0.0} & 0.8857±0.0126 (4)& 0.8016±0.0193 \\ 
   & \textbf{FuncVul}  & \textbf{0.9022±0.0157} & \textbf{0.8456±0.0212} & 0.9443±0.0282 & \textbf{0.8917±0.0178} (1) & \textbf{0.8947±0.0454} \\
  \midrule

&  CodeBERT & 0.8663±0.0176 & \underline{0.7856±0.0202} & 0.9512±0.0346 & 0.8602±0.0205 (4) & \underline{0.7545±0.0096}\\

% &   Gemini  & 0.8100±0.0232 & \underline{0.7901±0.0371} & 0.7963±0.1110 & 0.7963±0.0519 (4) \\ 
3 & CustomVulBERTa & \underline{0.8675±0.0114} & 0.7793±0.0210 & \underline{0.9682±0.0141} & \underline{0.8634±0.0155} (2) & 0.7544±0.0.021 \\
& BERT & 0.8054±0.0248  & 0.7762±0.0143 & 0.7764±0.0459 & 0.7758±0.0247 (5) & 0.6043±0.0521\\
& VUDENC & 0.7485±0.0199  & 0.7153±0.0258 & 0.6981±0.0306 & 0.7063±0.0244 (6) & 0.4867±0.0420\\
& LineVul & 0.8656±0.0121 & 0.7750±0.0240 & \textbf{0.9721±0.0121} & 0.8622±0.0156 (3) & 0.7527±0.0.0192 \\
&   \textbf{FuncVul}  & \textbf{0.8723±0.0114} & \textbf{0.7924±0.0245} & 0.9544±0.0183 & \textbf{0.8657±0.0174} (1) & \textbf{0.8825±0.0577} \\
\midrule

 & CodeBERT  & \underline{0.8735±0.0141} & \underline{0.7940±0.0281} & 0.9526±0.0254 & \underline{0.8654±0.0140} (2) & \underline{0.7602±0.0274}\\

 % & Gemini  & 0.7245±0.0581 & \underline{0.8005±0.0300} & 0.5290±0.1694 & 0.6204±0.1282 (6) \\ 
  4 & CustomVulBERTa & 0.8684±0.0133 & 0.7817±0.0206 & \underline{0.9600±0.0124} & 0.8616±0.0141 (3) & 0.7531±0.0.0239 \\ 
 & BERT & 0.80677±0.0158 & 0.7712±0.0285 & 0.7784±0.0288&0.7743±0.0198 (5) & 0.6058±0.0317\\

 & VUDENC & 0.7658±0.0155  & 0.7238±0.0253 & 0.7302±0.0401& 0.7263±0.0233 (6) & 0.5225±0.0321\\
 & LineVul & 0.8656±0.0163 & 0.7759±0.0264 & \textbf{0.9642±
0.0092} & 0.8596±0.0162 (4) & 0.7500±
0.0270 \\ 
& \textbf{FuncVul}  & \textbf{0.8797±0.0118} & \textbf{0.8077±0.0200} & 0.9426±0.0182 & \textbf{0.8698±0.0134} (1) & \textbf{0.7982±0.0470}\\ \midrule

&  CodeBERT & \underline{0.8914±0.0125 }& \underline{0.8914±0.0136} & 0.9148±0.0329 & \underline{0.9006±0.0131} (2) & \underline{0.8035±0.0253}\\

% &   Gemini  & 0.5607±0.0532 & 0.6256±0.0471 & 0.2164±0.1110 & 0.3215±0.0519 (6) \\ 
5 & CustomVulBERTa & 0.8905±0.0147 & 0.8530±0.0244 & \underline{0.9446±0.0208} & 0.8962±0.0130 (3) & 0.7860±0.0278\\
& BERT & 0.5897±0.0856  & 0.7902±0.1950 & 0.3860±0.3342 & 0.3997±0.3296 (6) & 0.2007±0.01448\\

&  VUDENC & 0.8034±0.0135 & 0.7996±0.0171 & 0.8097±0.0152 & 0.8045±0.0145 (5) & 0.6064±0.0269\\ 
&LineVul & 0.8509±0.0199 & 0.7895±0.0358 & \textbf{0.9596±0.0306} & 0.8655±0.0178 (4) & 0.7205±0.0340 \\
&   \textbf{FuncVul}  & \textbf{0.9004±0.0116} & \textbf{0.8934±0.0130} & 0.9096±0.0154 & \textbf{0.9013±0.0111} (1) & \textbf{0.9556±0.0041}\\
\midrule

 & CodeBERT  & \underline{0.8984±0.0071} &\underline{0.9007±0.0187} & 0.9330±0.0207 & \underline{0.9155±0.0073} (2) & \underline{0.8377±0.0137}\\
 
 % & Gemini  & 0.5353±0.0385 & 0.5352±0.0360 & 0.5162±0.0269 & 0.5255±0.0405 (6) \\ 
 6 & CustomVulBERTa & 0.8898±0.0170 & 0.8434±0.0416 & \textbf{0.9614±0.0269} & 0.8975±0.0124 (3) & 0.7900±0.0242\\ 
 & BERT & 0.7115±0.0839 & 0.7069±0.0959 & 0.7946±0.1116&0.7371±0.0258 (5) & 0.4465±0.1162\\
 &  VUDENC & 0.8290±0.0074 & 0.8196±0.0161 & 0.8443±0.0145 & 0.8316±0.0073 (4) & 0.6585±0.0144\\ 
 & LineVul & 0.7951±0.1676 & 0.6455±0.0.3612 & 0.7785±0.4355 & 0.7056±0.3944 (6) & 0.7570±0.006 \\
 & \textbf{FuncVul}  & \textbf{0.9184±0.0053} & \textbf{0.9056±0.0117} & \underline{0.9343±0.0116} & \textbf{0.9196±0.0057}  (1) & \textbf{0.9619±0.0031}\\

\bottomrule
\end{tabular}
}
\label{code_chunk_results}
\end{table}

\subsubsection{FuncVul Model Performance (RQ1): }

We compare our proposed FuncVul model against five baseline methods across six benchmark datasets. As shown in Table \ref{code_chunk_results}, FuncVul consistently outperforms all other models, ranking first in most evaluation metrics—including F1-Score, Accuracy, Precision, and MCC—across all datasets. Specifically, it achieves the highest score in 25 out of 30 cases and ranks second in two additional cases. The LineVul model also demonstrates strong performance, obtaining the highest Recall in four cases and second-best results in three others. CodeBERT and CustomVulBERTa exhibit competitive results in certain settings, with CodeBERT achieving the second-highest score in 18 cases. CustomVulBERTa achieves seven second-best results and one best-case performance. All experiments are conducted using an 80/20 train-test split, and results are reported as the average of five-fold cross-validation. To determine the best model, we rank F1-scores due to their balanced representation of Precision and Recall. Overall, Table \ref{code_chunk_results} confirms that FuncVul delivers the most consistent and superior performance, effectively answering RQ1.

% \begin{figure}[htp!]
%     \centering
%     ~\subfloat[Dataset 1]{ \includegraphics[width=0.49\linewidth]{Figure/D_3_code_chunks_FP_FN.png}}  
%      ~\subfloat[Dataset 2]{ \includegraphics[width=0.49\linewidth]{Figure/D_4_code_chunks_FP_FN.png}} 
%     \caption{FP and FN number for five fine-tune models on code chunks datasets.} 
%     \label{fig:Code_Chunks_FP_FN}
% \end{figure}
\subsubsection{Code Chunks vs Full Function based Results Analysis (RQ2):}
\begin{figure}[htp!]
    \centering
    ~\subfloat[Dataset 1]{ \includegraphics[width=0.49\linewidth]{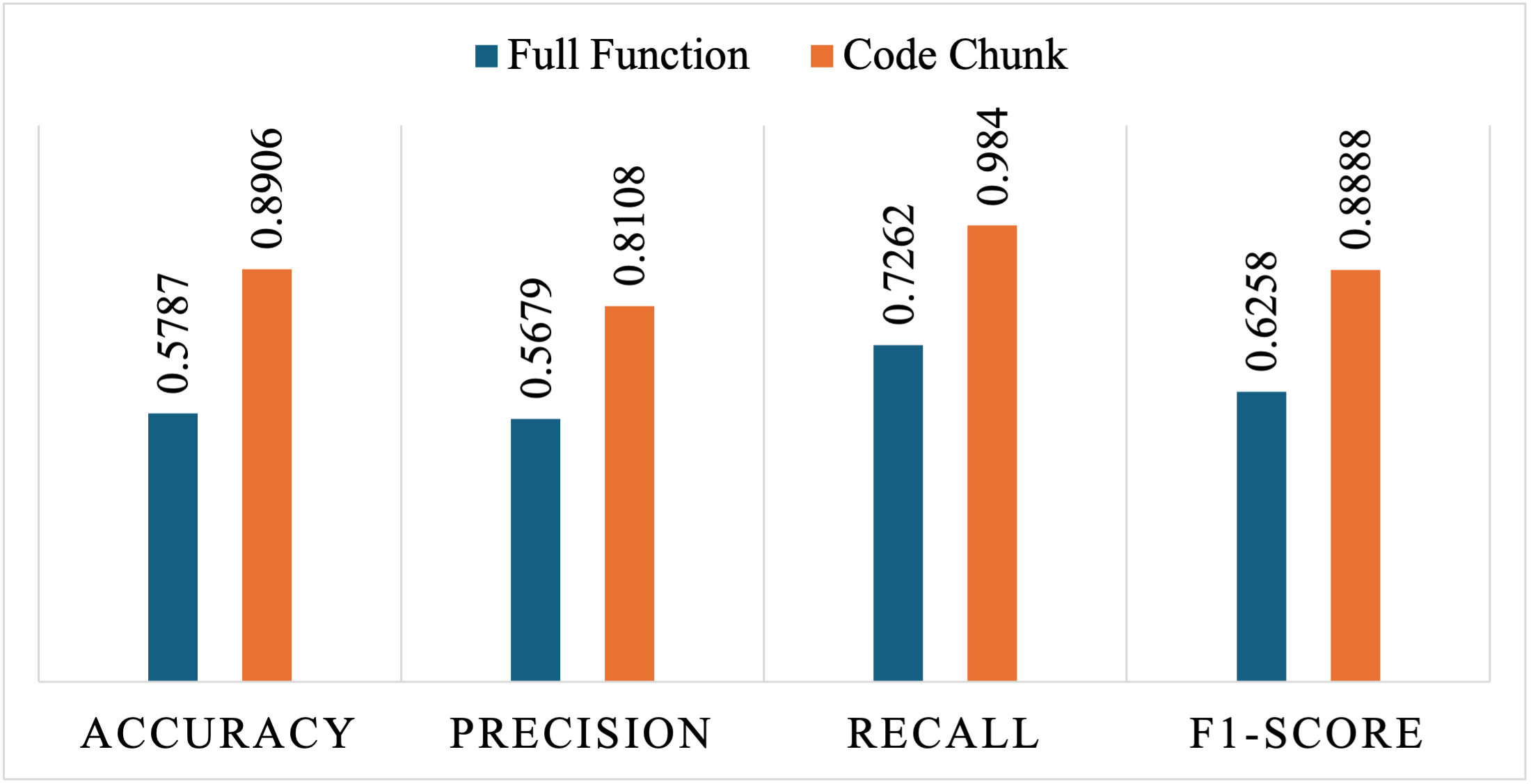}}  
     ~\subfloat[Dataset 2]{ \includegraphics[width=0.49\linewidth]{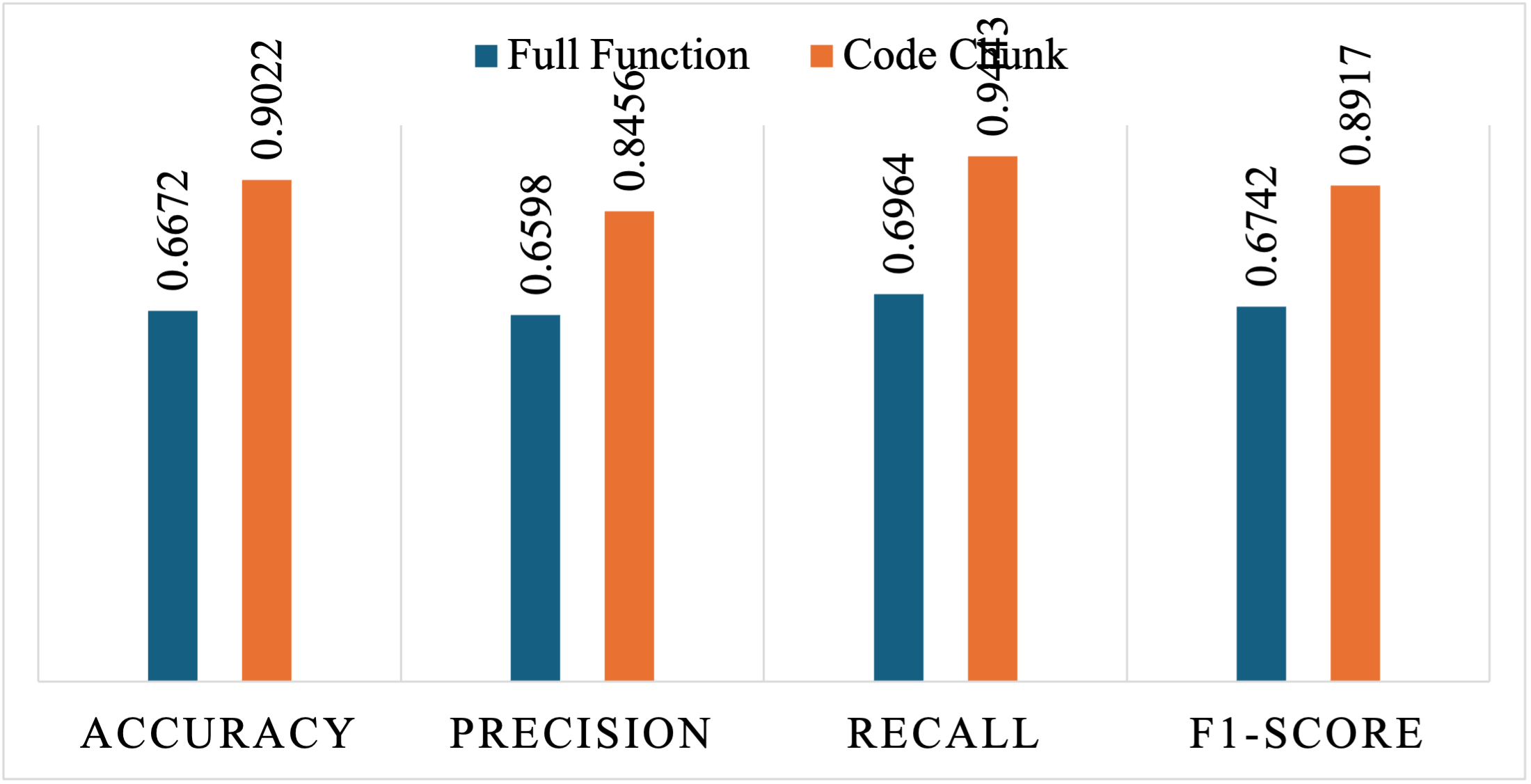}} 
    \caption{Comparison between our proposed code chunk based results with full function code based results.} 
    \label{fig:full_function_vs_code_chunk}
\end{figure}

Table \ref{fig:full_function_vs_code_chunk}(a) compares the performance of the Full Function and Code Chunk approaches on Dataset 1. The Code Chunk method significantly outperforms the Full Function approach across all metrics. It improves accuracy by ~53.9\%, precision by ~42.8\%, recall by ~35.5\%, and F1-score by ~42.0\%. We also gets the same kinds of results on Dataset 2 in figure \ref{fig:full_function_vs_code_chunk}(b). In this case our proposed code chunk based model improves accuracy by 35.22\%, precision by 28.16\%, recall by 35.59\% and F1-score by 32.26\%. 
These results highlight that the code chunk approach significantly enhances the model's capability for vulnerability detection than the full-function based approach. These findings effectively address our research question RQ2.

\subsubsection{Generic Code Chunks based Results Analysis (RQ3):}

\begin{figure}[htp!]
    \centering
    ~\subfloat{ \includegraphics[width=0.49\linewidth]{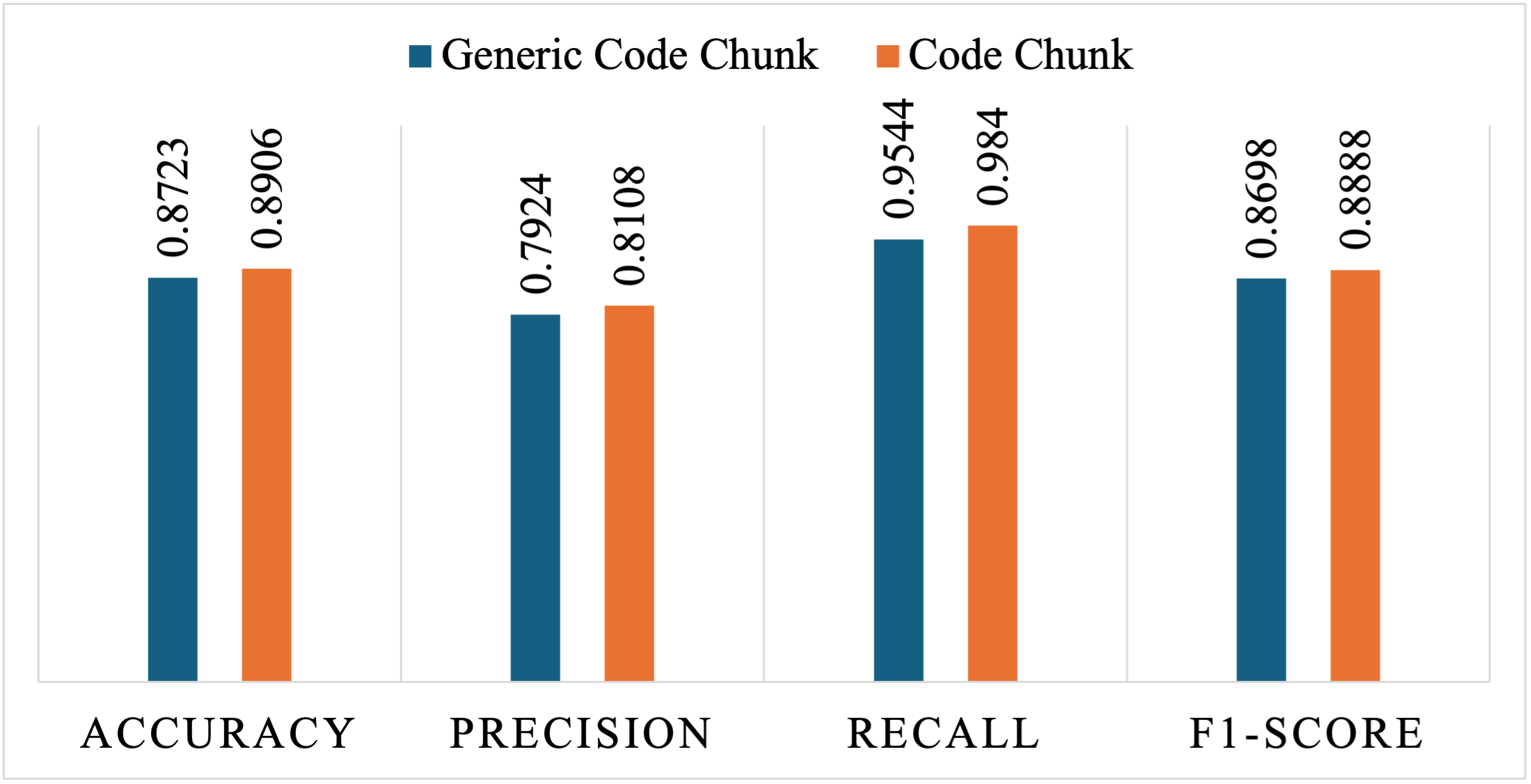}}  
     ~\subfloat{ \includegraphics[width=0.49\linewidth]{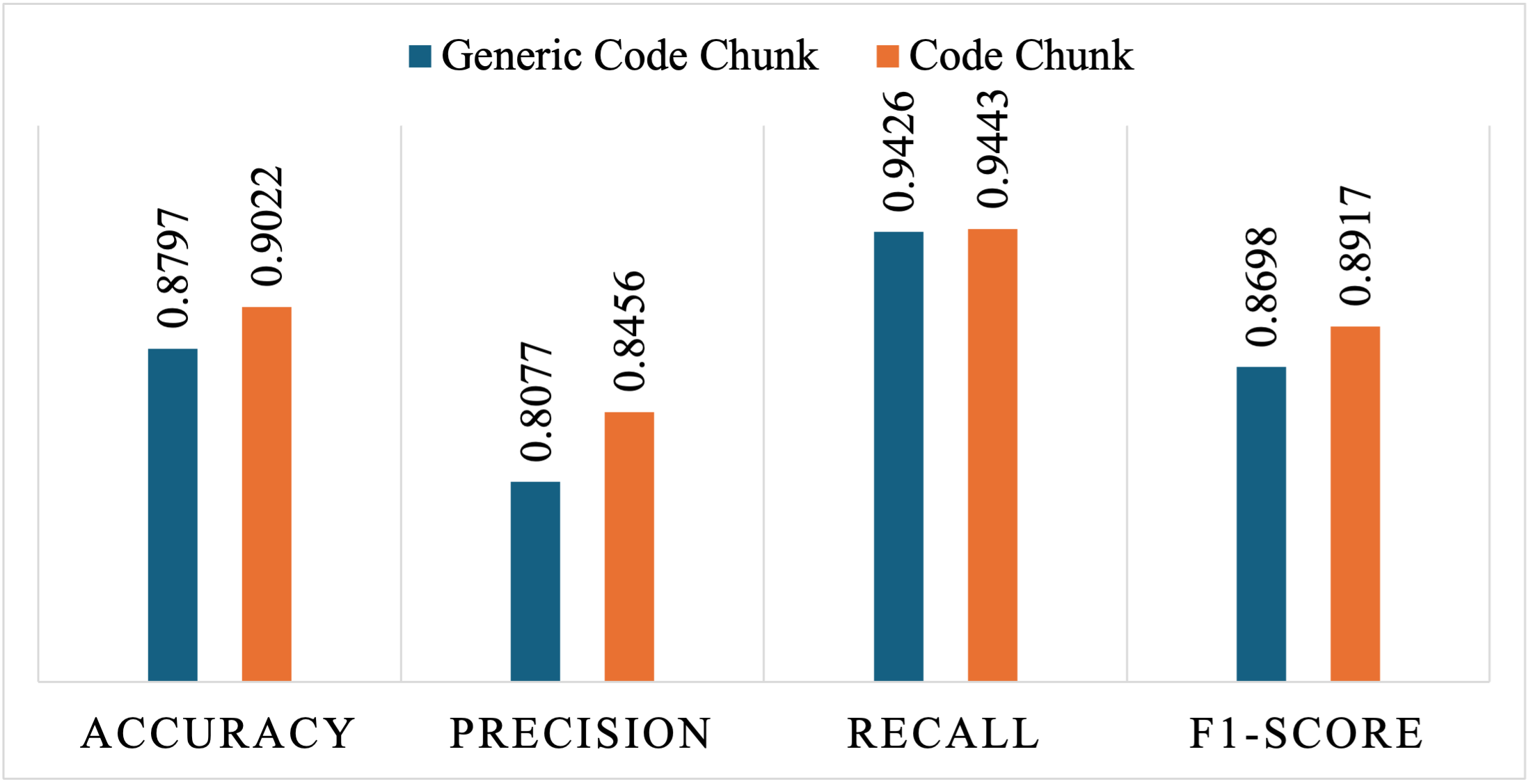}} 
    \caption{Comparison between our proposed code chunk based results with generic code chunk based results.} 
    \label{fig:generic_vs_code_chunk}
\end{figure}

In this work, we construct datasets based on code chunk and generic code chunk methodologies using the same data. Figure \ref{fig:generic_vs_code_chunk}(a) presents a comparative analysis between code chunk-based Dataset 1 and generic code chunk-based Dataset 3 results on FuncVul method. The results demonstrate that the code chunk based results consistently outperforms generic code across all evaluation metrics, achieving improvements of 2\% in Accuracy, 1.84\% in Precision, 2.96\% in Recall, and 1.9\% in F1-score. 
Similarly, Figure \ref{fig:generic_vs_code_chunk}(b) compares Code chunk-based Dataset 2 with generic Code Chunk-based Dataset 4, showing improvements of 2.25\% in Accuracy, 3.79\% in Precision, and 2.19\% in F1-score, with a modest 0.17\% gain in Recall. These findings underscore the superior effectiveness and adaptability of the Code Chunk approach, validating RQ3.

% \begin{figure}[htp!]
%     \centering 
%      ~\subfloat[Dataset 3]{ \includegraphics[width=0.49\linewidth]{Figure/D_3_generic_code_chunks_FP_FN.png}} 
%      ~\subfloat[Dataset 4]{ \includegraphics[width=0.49\linewidth]{Figure/D_4_generic_code_chunks_FP_FN.png}} 
%     \caption{FP and FN number for five fine-tune models on generic code chunks datasets.}
   
%     \label{fig:Code_Chunks_FP_FN}
% \end{figure}

% \begin{figure*}
%     \centering
   
% \end{figure*}

\subsubsection{Effectiveness of Identifying new CVEs and new Project CVEs vulnerabilities (RQ4): }

Previous results show vulnerability detection outcomes from five-fold cross-validation. However, Figure \ref{fig:label_data} shows many vulnerable packages remain undetected by LLM Detect, marked as 'Unknown'. To evaluate this unknown data, we created two test datasets. The first contains CVEs absent from training data, while the second includes code from different project IDs than the training data. We trained our proposed model FuncVul using 100\% data on Dataset 2. Table \ref{testdata} shows the two test case data in details.

\begin{table*}[htp!]
   \caption{New CVEs and new project ID-based test data. }
    \centering
     \def\arraystretch{1.0}\tabcolsep=8pt
     % \resizebox{0.48\textwidth}{!}{
    \begin{tabular}{ l l l c c}
       &  Case & Type & Vulnerable & Non-Vulnerable  \\ \hline 
       & Test Case 1 &   New CVSs & 1245 & 1753 \\ 
      &  Test Case 2 &  New Project ID & 179 & 280 \\ \hline 

      % Balance & Test Case 3 &   New CVSs & 1245 & 1245 \\ 
      % &  Test Case 4 &  New Project ID & 179 & 179 \\ \hline 
    \end{tabular}
     % }
    \label{testdata}
\end{table*}
Table \ref{tab:test_results_new_cves} shows that the FuncVul model achieves an accuracy of 81.95\% in Test Case 1 and 76.69\% in Test Case 2. This indicates that the model can correctly identify 81.95\%  of code chunks that were previously unexplored during dataset 2 construction (labeled as Unknown in Figure \ref{fig:label_data}). The FuncVul model shows strong capability in detecting vulnerabilities, particularly in recall 90.20\%. These results support the model's robustness in identifying unknown vulnerabilities, addressing the objectives of RQ4.

\begin{table*}[htp]
\caption{New CVEs and new project ID based prediction results for various model on Dataset 2.}
    \centering
    \def\arraystretch{1.0}\tabcolsep=4pt
   % \resizebox{0.98\textwidth}{!}{

    \begin{tabular}{l  l  l c c c c c c }
    \toprule
       & Case  & Model &  Accuracy  & Precision & Recall & F1-Score  & FP   & FN  \\ \midrule
        % & & CodeBERT& 0.8109 & 0.6883 & \underline{0.9952} & \underline{0.8138} & 561 & \underline{6} \\ 
        & Test Case 1 & FuncVul & 0.8195 & 0.7283 & 0.9020 & 0.8059 & 419 & 122 \\
        % & & CustomVulBERTa & \textbf{0.8306} & \underline{0.7102} & \textbf{1.0} & \textbf{0.8306} & \underline{508} & \textbf{0}\\
        % && BERT & 0.8182 & 0.7444 & 0.8562 & 0.7964 & 366 & 179 \\ 
        \midrule
        % & & CodeBERT & 0.7516 & 0.6125 & \underline{0.9888} & \underline{0.7564} & 112 & \underline{2} \\ 
      &  Test Case  2 & FuncVul & 0.7669 & 0.6552 & 0.8492 & 0.7397 & 80 & 27 \\
      % \multirow{-6}{2mm}{\rotatebox{90}{}} & & CustomVulBERTa & \textbf{0.7821} & \underline{0.6416} & \textbf{1.0} & \textbf{0.7817} &   \underline{100} & \textbf{0} \\ 
      % & & BERT & 0.7647 & 0.6749 & 0.7654 & 0.7173 & 66 & 42 \\ 
        \midrule
         % & & CodeBERT& 0.8378 & 0.7569 & 0.9952 & 0.8598 & 398 & 6  \\ 
        %  & Test Case 3 & FuncVul & 0.8329& 0.7925 & 0.9020& 0.8437 & 294 & 112 \\
        % %  & & CustomVulBERTa & \textbf{0.8558} & 0.7762 & \textbf{1.0} & 0.8740 & 359& 0 \\
        % % && BERT & 0.8257 & 0.8070 & 0.8562 & 0.8309 & 255 & 179 \\ 
        %  \midrule
        %  % & & CodeBERT & \underline{0.7989} & 0.7166 & \underline{0.9888} & \underline{0.8310} & 70 & \underline{2} \\ 
        %   & Test Case 4 & FuncVul & 0.7961 & 0.7677& 0.8492 & 0.8064 & 46 & 27 \\ 
        %   % & & CustomVulBERTa &  \textbf{0.8268} & \underline{0.7427} & \textbf{1.0} & \textbf{0.8524} & 62 & \textbf{0} \\
        %   % && BERT & 0.7709 & 0.7740 & 0.7654 & 0.7697 & 40 & 42 \\ 
          \bottomrule
         
    \end{tabular}
    % }
     \label{tab:test_results_new_cves}
\end{table*}

\subsubsection{Impact of Line Numbers to Create Code Chunks (RQ5):}

\begin{figure}[htp!]
    \centering
    ~\subfloat[Accuracy]{ \includegraphics[width=0.24\linewidth]{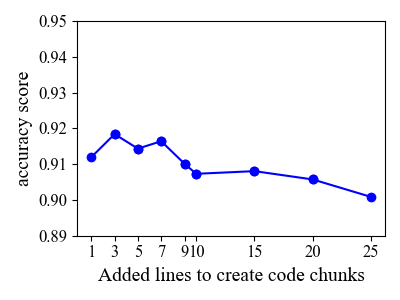}}
    ~\subfloat[Precision]{ \includegraphics[width=0.24\linewidth]{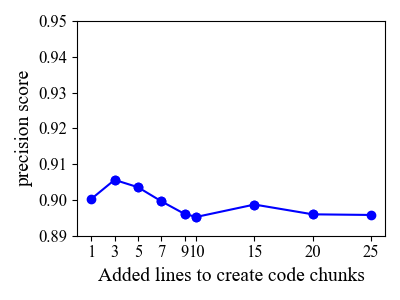}} 
    ~\subfloat[Recall]{ \includegraphics[width=0.24\linewidth]{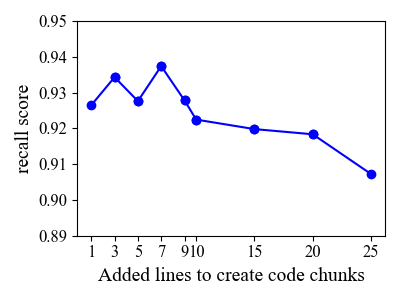}} 
     ~\subfloat[F1 Score]{ \includegraphics[width=0.24\linewidth]{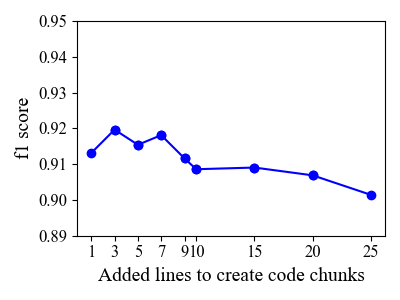}} 
    \caption{Impacts of code chunk length on dataset 6.} 
    \label{fig:Code_Chunks_length}
\end{figure}

% In this research, our primary objective is to generate concise code chunks that effectively highlight vulnerable patterns. To achieve this, we adopted ``3-line extended based code chunk", which involves including three lines before and three lines after a detected vulnerable line to form the code chunks. Now, one question may arise: why did we choose the 3-line extended based code chunk? To validate this choice, we conducted experiments using code chunks extended by different numbers of lines—namely 1, 5, 7, 9, 10, 15, 20, and 25—and assessed their performance with our proposed FuncVul model. The results demonstrate that the 3-line extended based code chunk outperforms other configurations in most metrics. Specifically, Figure \ref{fig:Code_Chunks_length} (a) shows that the accuracy of the 3-line extended code chunks on Dataset 6 is the highest among all configurations. Similarly, Figure \ref{fig:Code_Chunks_length} (b) highlights superior precision for the 3-line extended strategy.
Our primary objective is to generate concise code chunks highlighting vulnerable patterns. We adopted "3-line extended based code chunk", including three lines before and after a detected vulnerable line. To validate this approach, we tested code chunks extended by different line numbers—1, 5, 7, 9, 10, 15, 20, and 25—and assessed their performance with our FuncVul model. Results show the 3-line extended code chunk outperforms other configurations in most metrics. Figure \ref{fig:Code_Chunks_length} (a) shows highest accuracy for 3-line extended chunks on Dataset 6, while Figure \ref{fig:Code_Chunks_length} (b) demonstrates superior precision for this strategy.
Although the recall score, shown in Figure \ref{fig:Code_Chunks_length} (c), is highest for 7-line extended code chunks, the 3-line extended approach ranks second. Figure \ref{fig:Code_Chunks_length} (d) shows the F1 score, balancing recall and precision, confirming that 3-line chunks provide the best performance. Larger chunks make it difficult to identify vulnerable code effectively, while the 3-line strategy's average chunk length of 6.2 lines provides optimal balance between context and conciseness for detecting vulnerable patterns.

% Although the recall score, as depicted in Figure \ref{fig:Code_Chunks_length} (c), is highest for the 7-line extended code chunks, the 3-line extended approach ranks a close second. Finally, Figure \ref{fig:Code_Chunks_length} (d) illustrates the F1 score, which balances recall and precision, confirming that the 3-line code chunks provide the best overall performance.

% The main reason for this superior performance lies in the length of the generated code chunks. Larger chunks makes it difficult to pinpoint one or two lines of vulnerable code effectively. In contrast, the 3-line extended strategy yields an average chunk length of 6.2 lines, providing an optimal balance of context and conciseness, thereby enhancing the detection of vulnerable code patterns.

\subsection{Detection of Multiple Vulnerabilities Within Function Code (RQ6):}
In this work, we use code chunk based data for model built that can split one single function in multiple code chunk. Thus, the proposed model is capable of detecting multiple vulnerabilities within a single function's code. By analyzing smaller, context-rich segments, the model effectively captures various vulnerability patterns, enabling comprehensive detection across different parts of the function.
% \subsection{Discussion}

\section{Conclusion}
\label{conclusion}
In this paper, we present FuncVul, a novel model for function-level vulnerability detection that leverages function code chunks and the pre-trained GraphCodeBERT model. Unlike existing approaches, FuncVul not only identifies whether a function is vulnerable but also detects the specific number of vulnerable code chunks, significantly reducing the time required by developers or experts to address vulnerabilities. Experimental results demonstrate that FuncVul outperforms baseline models on both code chunk and generic code chunk datasets. Additionally, our analysis reveals that datasets based on three-line code chunks from large language models yield higher accuracy and F1-scores compared to datasets where patch information is derived by removing lines of code. Furthermore, we demonstrate that our dataset can be generalized using large language models, resulting in enhanced model performance.

% In this study, we focus solely on vulnerability detection strategies. We acknowledge that not all vulnerabilities pose the same level of risk; hence, we did not consider multi-class vulnerability detection. In future work, we aim to extend this research to include multi-class vulnerability detection to better address varying levels of risk. We also focus on identifying vulnerabilities in C/C++ and Python code. In future work, we plan to extend our model to detect vulnerabilities in other programming languages.
This study focuses on vulnerability detection strategies, without considering multi-class detection since vulnerabilities have varying risk levels. Future work will extend to multi-class vulnerability detection to address risk variations. Currently, we identify vulnerabilities in C/C++ and Python code, with plans to expand to other programming languages.

 % \section*{Acknowledgement}

\bibliographystyle{splncs04}
\bibliography{myb}

\appendix

% \newpage
\section{Appendix}

\subsection{Generic Code Conversion LLM Prompt}
\label{generic_prompts}
Generic code chunks converts code chunks in generic format. Here, we use the following LLM to convert the generic code chunk as described in Table \ref{tab:generic_prompt}. Table \ref{tab:generic_prompt} also shows the one example of code chunk and converted generic code chunk based on the proposed LLM generic prompt.

\begin{table}[htp!]
   \captionsetup{position=top}  % Forces the caption to appear at the top
     \caption{Prompt and Example for Transforming Code Chunks into Generic Code Chunk}  
     \label{tab:generic_prompt}
     \centering
    \begin{tabular}{p{1\textwidth}}
     \toprule
        \textbf{Description of Generic Prompt}\\ \midrule
           Here is the function code chunk: $\{code\_chunk\}$
                
            Please convert the code chunk by renaming functions to $F_1, F_2, ..., F_N$ and variables to $v_1, v_2, ..., v_n$. 
                
            Return the converted code in a variable named \emph{generic\_code}. 
                \\ \midrule \midrule
            Example \\
        \midrule \midrule 
       
                Code Chunk  \\ \midrule
                
                \hspace{0.5cm} goto trunc;
                
                if (length $<$ alen)
                
                \hspace{0.5cm}    goto trunc;
                    
                if (!bgp\_attr\_print(ndo, atype, p, alen))
                
                \hspace{0.5cm}    goto trunc;
                    
                p += alen;
                
                len -= alen; \\ \midrule 

                 Generic Code Chunk \\ \midrule

                \hspace{0.5cm}goto F1;
                
                if (v1 $<$ v2)
                
                \hspace{0.5cm}    goto F1;
                
                if (!F2(v3, v4, v5, v2))
                
                \hspace{0.5cm}    goto F1;
                
                v5 += v2;
                
                v6 -= v2;
               \\ \bottomrule

    \end{tabular}
    
\end{table}

\subsection{Vulnerable Samples detection LLM Prompts }
In this paper, we construct six datasets using two distinct LLM prompts. Figure \ref{prompt_1_2} provides a detailed overview of these prompts—one utilizing only the code and the other combining a description with the code.
\label{vulnerable_samples_prompts}

\begin{table}[ht!]
    \centering
    \caption{LLM prompts for detecting vulnerable samples with different input settings.}
    \label{prompt_1_2}
    \begin{tabular}{p{0.25\textwidth} | p{0.7\textwidth}}
    \hline
    \textbf{Prompt Type} & \textbf{Input Context} \\ \hline
    Code Only & Given the following function code: \{\texttt{code}\} \\ \hline
    Code + Description & Given the following function code: \{\texttt{code}\} \\
    & And the associated CVE description: \{\texttt{desc}\} \\ \hline
    \multicolumn{2}{p{0.95\textwidth}}{
    \textbf{Task:} Extract the following information:
    \begin{itemize}
        \item  [1.] Identify the lines of code that contain vulnerabilities. Return these lines in a list of string named as \emph{line\_code}. If no vulnerable lines are found, return ['None']. Ensure the list is formatted with items separated by commas and enclosed in square brackets.
        
         \item [2.] Determine the line numbers of vulnerable code. Return these line numbers in a list of integer named as \emph{vul\_lines}. If no such lines exist, return ['None'].

        \item [3.] List the affected vulnerability categories. Return these in a list of string named as \emph{vul\_category}. If no categories are affected, return ['None'].
                    
    \end{itemize}
     Please provide the output in three keys as dictionary format: \emph{line\_code}, \emph{vul\_lines}, and \emph{vul\_category}. Do not need an explanation. 
    } \\
    \hline
    \end{tabular}
\end{table}

\subsection{Details of Baselines}
\label{baselines}

We compare our proposed FuncVul model with five baselines that are as follows. 
\begin{itemize}
    \item \textbf{CodeBERT} \cite{feng2020codebert}: CodeBERT is a pre-trained model for understanding and generating both natural language and programming code.

    % \item \textbf{DeekSeek} \cite{guo2024deepseek}: It is a deep learning language model with 7 billion parameters, optimized for code vulnerability detection and natural language processing.
    % \item \textbf{GPT-3.5 Turbo} \cite{openai2024gpt35turbo}: GPT-3.5 Turbo is a faster, cost-effective version of OpenAI's GPT-3.5 model, designed for efficient and high-performance natural language processing.
    
    % \item \textbf{Gemini-1.5 Pro} \cite{google2024gemini15pro}: It is an advanced AI language model developed by Google DeepMind, designed for high-performance natural language processing and understanding.

    % \item \textbf{CustomVulBERTa}: CustomVulBERTa is a fine-tuned version of the VulBERTa \cite{hanif2022vulberta} model, specifically adapted for detecting security vulnerabilities in source code. VulBERTa is a pre-trained RoBERTa-based model that utilises a tailored tokenization pipeline for real-world C/C++ code from open-source projects. In our baseline, we fine-tune CustomVulBERTa on our dataset to ensure a reliable benchmark for result comparison. 

    \item \textbf{CustomVulBERTa}: CustomVulBERTa is a fine-tuned version of VulBERTa \cite{hanif2022vulberta}, a RoBERTa-based model pre-trained on real-world C/C++ code, adapted to detect security vulnerabilities and used as a baseline in our experiments.
    
    \item \textbf{BERT}: BERT \cite{kenton2019bert} is a pretrained deep bidirectional transformer model that uses masked language modeling to capture context from both directions, distinguishing it from unidirectional or shallow concatenation-based approaches.

    \item \textbf{VUDENC}: VUDENC \cite{wartschinski2022vudenc} is a deep learning tool designed to detect vulnerabilities in real-world Python code. It uses a word2vec model to generate vector representations of semantically similar code tokens and employs LSTM networks to classify vulnerable code sequences.
    \item \textbf{LineVul}: LineVul \cite{fu2022linevul} is a Transformer-based fine-grained line-level vulnerability prediction model.

    % \item \textbf{FuncVul} : We propose a code chunk-based fine-tuned version of GraphCodeBERT \cite{guo2020graphcodebert}, which is capable of identifying multiple vulnerabilities and effectively detecting function-level vulnerabilities.
    
\end{itemize}

\subsection{Details of Evaluation Metrics}
\label{evaluatoin_metric_details}
To evaluate the performance of our proposed model, we employ widely used prediction evaluation metrics, as outlined below.
\begin{itemize}

\item \textbf{Accuracy:} It measures the overall correctness of a model in predicting both code vulnerabilities and non-vulnerabilities. $Accuracy = \frac{TP + TN}{ TP + FP + TN + FN}$

\item \textbf{Precision:} It measures the proportion of true code vulnerabilities to the total number of code vulnerabilities that have been predicted as vulnerabilities by the model: $Precision = \frac{TP}{TP + FP }$

\item \textbf{Recall:} It measures the proportion of true vulnerabilities detected by a model to the total number of code vulnerabilities in the dataset: $Recall = \frac{TP} {TP + FN}$
    
\item \textbf{F1-Score:} It is the harmonic mean of precision and recall: 

{\centering $F1\text{-}Score = \frac{2 * Precision * Recall} {Precision + Recall}$}

% \item \textbf{FP Number:} False positive (FP) occurs when the model incorrectly identifies a code chunk as vulnerable when, in reality, it is non-vulnerable. 

% \item \textbf{FN Number:} False negative (FN) means the model incorrectly identifies a code chunk as non-vulnerable when, in reality, it is vulnerable. 

\item \textbf{Matthews Correlation Coefficient (MCC):} MCC is a robust metric that reflects balanced performance across all confusion matrix categories and is particularly effective for evaluating models on imbalanced datasets.  

{\centering $MCC = \frac{TP \cdot TN - FP \cdot FN}{\sqrt{(TP + FP)(TP + FN)(TN + FP)(TN + FN)}}$}

\end{itemize}
\end{document}